\documentclass[twocolumn,preprintnumbers,amsmath,amssymb,aps]{revtex4}
\usepackage{graphicx}% Include figure files
\usepackage{dcolumn}% Align table columns on decimal point
\usepackage{bm}% bold math

\begin{document}

%\preprint{astro-ph/}

\title{On the nature of dark matter\footnote{Honorable Mention in the 2004 Awards for Essays on Gravitation (Gravity Research Foundation)}}% Force line breaks with \\

\author{Tonatiuh Matos}
 \email{tmatos@fis.cinvestav.mx}
\affiliation{%
Departamento de F{\'\i}sica, Centro de Investigaci\'on y de
Estudios Avanzados del IPN, A.P. 14-740, 07000 M\'exico D.F., M\'exico.
}%
\author{L. Arturo Ure\~na-L\'opez}
 \email{lurena@fisica.ugto.mx}
\affiliation{%
Instituto de F\'isica de la Universidad de Guanajuato, A.P. 150,
37150, Le\'on, Guanajuato, M\'exico.
}%


\begin{abstract}
Dark matter in the universe seems to be one of the most important puzzles science has to face in this moment. In this essay we point out that dark matter could be a spin-0 fundamental interaction of Nature rather than a simple particle. From this hypothesis follows that dark matter behaves just as standard cold dark matter at cosmological level while still in good agreement with observations at galactic scales. This new interaction could be one of the scalar fields predicted by higher-dimensional theories.
\end{abstract}

\maketitle
For many years, it seemed that our deepest understanding of Nature
would come from the study of the tiniest constituents of matter.
That feeling changed drastically once the scientific community was
able to see the small imperfections of the radiation sea that
fills all of the observable universe. If we
agree with the most recent cosmological observations, we should
instead turn our attention to the largeness if we want to complete
our knowledge of Nature.

To put it in numbers, there is increasing evidence for the
existence of exotic matter which would account for 96\% of the
total material content of the universe\cite{cmbr,scp}. The simplest picture makes
the following division. 73\% of the cosmic matter is
gravitationally repulsive and has been labelled as \textit{dark
energy}. On the other hand, 23\% of the cosmic matter is
gravitationally attractive and has been dubbed \textit{dark
matter}. The adjective \textit{dark} means that they do not emit
electromagnetic waves nor they interact with the particles we know
from the Standard Model of Particles (SMP). But, their presence
was detected by the important gravitational effects they have
imprinted on the evolution history of the cosmos.

Surprisingly, it has been possible to assemble a simple and successful
model\cite{basecosmo}, the so called Lambda Cold Dark Matter (LCDM), in which the
dark energy is represented by a cosmological constant ($\Lambda$),
and the dark matter is \emph{cold}. There is a plethora of specific
candidates that can play the role of dark energy and dark matter,
being the particles from the minimal supersymmetric extension of
the SMP the most popular. The LCDM model is very successful at
cosmological level: it accounts for the large scale structure and the accelerated expansion of the universe, for the micro-Kelvin
fluctuations of the Cosmic Microwave Background, etc. Moreover,
the aforementioned coldness of dark matter predicts well, in
accord with cosmological observations, the evolution of small
perturbations on a homogeneous and isotropic universe.
However, the model fails at galactic level. It predicts cusp
density profiles in galaxies, but new high resolution observations
indicate that density profiles in the center of galaxies are
almost flat \cite{deBlok,Rubin1,Rubin2,Rubin3,Blais,Bolato,Simon}. In addition, it seems that a new thorn has beed added to the CDM model: old galaxies were asambled at a higher redshift than expected\cite{gemini04a}.

As it can be inferred from the last paragraph, the case of dark
matter is of particular importance due to its key role in the
formation of galaxies, the unit cells of cosmological structure.
It is not surprising then that the search for dark matter
particles has become one of the most important challenges of
science \cite{Flores,dmneut,dmdirect,dmsearch}.

There are many theoretical proposals that go
beyond the SMP; among which the most successful are the
superstring theory and the brane models. A common feature of all of these super-theories is the existence of \textit{scalar fields}: the so called dilatons, radions, etc.,
which are key elements by themselves. The presence of scalar
fields in these models is surprising, to say the least, since no
fundamental scalar field has ever been detected, nor even the
so-needed Higgs particle of the SMP. To our knowledge,
there are only four fundamental interactions in Nature: gravitational,
electromagnetic, strong and weak. The first two are of long range,
while the other two are of short range, being the gravitational one
the weakest of all of them. Gravitation is a spin-$2$ interaction, and
the others are spin-$1$.

As there is not evidence for the existence of a fundamental scalar
field nor of a spin-$0$ interaction, it seems that our
super-theories have to have an internal mechanism to suppress
their own scalar fields. But, the evidence is not at all
conclusive and we should ask: \textit{did Nature
forget the most simple interaction?}

On one hand, we have to face the theoretical existence of
fundamental scalar fields; on the other hand, we have to face
the problem of the missing matter in the universe. A deal seems
possible: could it be possible that Nature did include the spin-$0$
interaction and made it the dominant one as part of what we called
dark energy and dark matter?

Let us construct a model of dark matter using a scalar field,
having in mind that, more than a particle, \emph{dark matter could
be a new fundamental interaction}; we shall call it the Scalar
Field Dark Matter (SFDM) model. With the theoretical and
observational evidence we have at hand, let us make the following
reasoning.

First of all, because of the success of the LCDM model, we should
impose the condition that the scalar field should mimic the
behavior of the CDM model at cosmological scales. Second, SFDM
should provide us of a new perspective with respect to the
formation and the structure of galaxies, that could make the
theoretical expectation compatible with actual observations.

Third, from theoretical reasoning, the scalar field must be
stabilized in some manner. One possibility is to endow the scalar
field to a scalar potential with a minimum. Fourth, the SFDM
models should not alter the successful early universe of the LCDM
model, like for instance, the epoch of nucleosynthesis. If possible, the scalar field should be subdominant at those
epochs.

One of the simplest scalar potentials that would fulfill the above
requirements is a $\cosh$ potential\cite{sahni01,MUL00a}.
Therefore, we propose the following effective Lagrangian for a L-SFDM model,
\begin{eqnarray}
{\cal L}_\textrm{L-SFDM}&=& {\cal L}_\textrm{GR}+{\cal L}_\textrm{B} + {\cal L}_\Lambda - \sqrt{-g} \left[ \Phi^{,\mu} \Phi_{,\mu} + 2 V(\Phi) \right] \, , \nonumber \\
V(\Phi) &=& \frac{m^2_\Phi}{8\pi G \lambda^2} \left[ \cosh\left(\sqrt{8\pi G} \lambda \Phi \right)-1 \right] \label{lagrangian}
\end{eqnarray}
where $\lambda$ and $m_\Phi$ are the free parameters of the model;
the latter is recognized as the mass of the scalar field. The
values of $\lambda$ and $m_\Phi$ can be fixed by imposing that the
SFDM model should reproduce the success of CDM at cosmological
scales\cite{MUL00a,MUL01a}, and then
\begin{eqnarray}
\lambda \sim 20 \, , && m_\Phi \sim 10^{-23} \, \textrm{eV} \, . \label{mass}
\end{eqnarray}

A brief description of the cosmological model that arises from~(\ref{lagrangian}) is as follows.
After inflation, the universe is dominated by the radiation
component, and the scalar potential is exponential (the scalar
field is far from the minimum). The large value of $\lambda$ makes
the scalar matter subdominant and behave as part of the radiation
fluid. It is in this epoch too that the field approaches the
minimum of the potential, ceases to follow the radiation fluid,
and begins to behave as a dust fluid. From this time on, the
evolution of the homogeneous and isotropic universe proceeds
exactly as in the LCDM model.

The latter is also true for the development of scalar
fluctuations, which grow as in the CDM case, except that all
structure on scales smaller than $\lambda_\Phi=m^{-1}_\Phi \sim 10
\, \textrm{pc}$ is suppressed. This eventually prevents, once a
scalar fluctuation collapses and forms a gravitationally bound
object, the formation of density profile with a central cusp. This
is confirmed by the numerical simulations for the collapse of a
single scalar fluctuation, which ends as a self-gravitating scalar
object called
\emph{oscillaton}\cite{seidel91,phi2002c,phi2003,don2003}.
The energy density profile of oscillatons is regular everywhere.

Oscillatons are stable scalar configurations if their mass is
below the critical value\cite{phi2003}
\begin{equation}
M_\textrm{crit}\sim 0.6\frac{m_\textrm{Pl}^2}{m_{\Phi}} \, .
\end{equation}
For the scalar mass value~(\ref{mass}), we get $M_\textrm{crit}
\sim 10^{12}M_\odot$, which is roughly the mass content of a
typical galaxy. Also, numerical studies within the weak-gravity
regime have shown that a galaxy scalar fluctuation rapidly
virializes\cite{laul2003a}, which seems to be in accord with
observations of the oldest galaxies in the
universe\cite{gemini04a}.

In this essay, we have made a r\'esum\'e of a promising SFDM
scenario, which offers the same results as the concordance LCDM
model at large scales, and also new vistas for galaxy formation
that seem to be in accord with actual observations. If realistic,
this model would give more than a simple answer to a major puzzle
of current Cosmology. It would make us think of a largely
predicted (by our supertheories) simple spin-$0$, long range
interaction of prime influence in the evolution of the universe.

Last but not least, there may be a connection between the dark
matter scalar field and the inflaton field. Such a connection was
explored in\cite{LMUL02} within the braneworld scenario for the
second Randall-Sundrum model. Though the predicted inflationary
quantities (spectral index, the amplitude of primordial
gravitational waves) are at variance with observations, the idea
is compelling: could a dark matter scalar field be the only
responsible for the \emph{whole} process of structure formation in
the universe, from primordial perturbations to galaxies? The
answer could be yes.

\begin{acknowledgments}
TM acknowledges partial financial support by CONACyT M\'exico,
under grants 32138-E and 34407-E. This work is also partially
supported by PROMEP and by projects of the Universidad de
Guanajuato.
\end{acknowledgments}

\bibliography{award}% Produces the bibliography via BibTeX.

\end{document}